\documentclass[prl,aps,floatfix,twocolumn]{revtex4}

\usepackage{graphicx}
\graphicspath{{figures/}}

\usepackage{units,xspace}
\newcommand{\micron}{\ensuremath{\unit{\mu m}}\xspace}

\usepackage{amsmath}
\usepackage{amsbsy}
\renewcommand{\vec}[1]{\boldsymbol{#1}}

\begin{document}

\title{Multidimensional
  optical fractionation with holographic verification}

\author{Ke Xiao}

\author{David G. Grier}

\affiliation{Department of Physics and Center for Soft Matter Research,
New York University, New York, NY 10003}

\date{\today}

\begin{abstract}
The trajectories of
colloidal particles driven through 
a periodic potential energy landscape can become kinetically
locked in to directions dictated by the landscape's symmetries.
When the landscape is realized with forces exerted by
a structured light field,
the path a given particle follows
has been predicted to depend
exquisitely sensitively on such properties
as the particle's size and refractive index
These predictions, however, have not been tested experimentally.
Here, we describe measurements 
of colloidal silica spheres'
transport through arrays of holographic optical traps that use
holographic video microscopy to track individual spheres'
motions in three dimensions and simultaneously
to measure each sphere's radius and refractive index
with part-per-thousand resolution.
These measurements confirm previously untested
predictions for the threshold of kinetically locked-in transport,
and demonstrate the ability of optical fractionation to sort 
colloidal spheres with
part-per-thousand resolution on multiple characteristics simultaneously.
\end{abstract}

\pacs{05.60.Cd, 82.70.Dd, 42.40.-i, 87.80.Cc}

\maketitle

Optical fractionation is akin to sifting particulate 
matter with a sieve made of light \cite{korda02b,macdonald03,ladavac04}.
Colloidal particles flowing with or through a viscous medium
such as water are deflected into different directions by
their differing interactions with a structured light field
\cite{korda02b}.
Unlike photophoresis \cite{hirai96}, in which particles are deflected down
the optical axis by radiation pressure,
optical fractionation
relies on optical forces directed transverse to the optical axis
\cite{ashkin86,roichman08}
and on the symmetries and periodicities of the structured light field
to select different fractions from an initially heterogeneous sample
\cite{macdonald03,ladavac04,pelton04a,gopinathan04,sancho05,gleeson06,lacasta06}.

\begin{figure}[!t]
  \centering
  \includegraphics[width=\columnwidth]{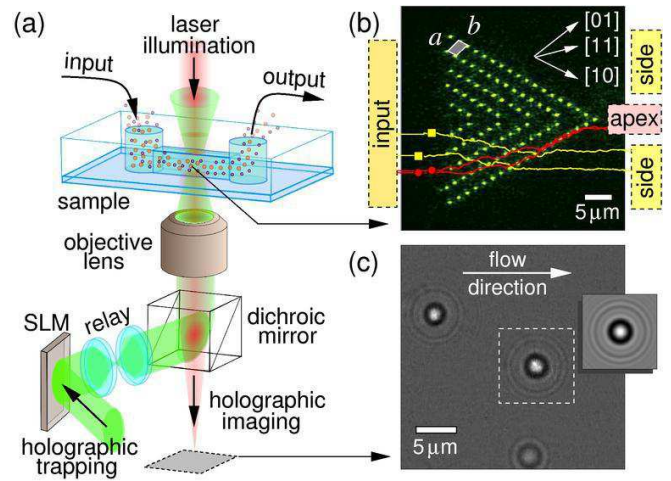}
  \caption{(color online)(a) Combined holographic optical trapping and 
    holographic video microscopy system for tracking and characterizing
    colloidal spheres during optical fractionation.
    The objective lens focuses computer-generated holograms projected
    with a spatial light modulator (SLM)
    into the sample.  It also collects
    holographic images of the sample that are recorded
    with a video camera.
    Trapping and imaging light are
    separated with a dichroic mirror.
    (b) Image of 103
    holographic optical tweezers in the microscope's focal plane.
    The traps are arranged in two domains of oblique
    lattice points with $a = 3.27~\micron$ and $b = 2.025~\micron$.
    One unit cell is highlighted.
    The domains are oppositely inclined so that
    $\theta_{[10]} = \pm 30^\circ$, $\theta_{[01]} = \mp 38^\circ$ and
    $\theta_{[11]} = \mp 13^\circ$ with respect to the flow direction.
    Dashed boxes indicate input and output regions.
    Traces show trajectories for four representative particles
    measured at 1/30~\unit{s} intervals.
    (c) Normalized holographic image of 
    1.5~\micron diameter silica spheres in the same field of view.
    Inset: Fit to the Lorenz-Mie theory of the 
    outlined sphere's holographic image.}
  \label{fig:schematic}
\end{figure}

Pioneering studies
\cite{korda02b,macdonald03,ladavac04,pelton04a,gopinathan04,gleeson06}
have suggested
that optical fractionation may
offer unparalleled selectivity for particle size and refractive index. 
Lacking sufficiently precise methods to test these predictions
directly, experimental studies
either have focused on sorting polydisperse samples
of particles with easily distinguishable properties 
\cite{macdonald03,ladavac04,milne07} 
or else have used indirect approaches to assess the method's finesse
\cite{roichman07a}.

This Letter describes optical fractionation measurements on
nominally monodisperse colloidal samples whose 
performance is verified directly
using the high-resolution single-particle 
characterization and tracking capabilities 
\cite{LMM}
of holographic video microscopy 
\cite{HVM}.
These measurements demonstrate optical fractionation on the basis
of size and refractive index with part-per-thousand resolution
and agree quantitatively with predictions 
derived from limiting arguments for colloidal transport through 
optical trap arrays \cite{ladavac04,pelton04a}.
They thus cast new light on the robustness of kinetically 
locked-in transport against
the inevitable disorder in practical
implementations.

Our experimental system is depicted schematically in
Fig.~\ref{fig:schematic}.
Aqueous dispersions of colloidal silica spheres are
driven by a syringe pump (Harvard Apparatus PHD 2000)
at a constant flow rate of $4~\unit{\mu l / h}$
through a rectangular microfluidic channel
formed by bonding a poly(dimethylsiloxane)
casting to a glass microscope cover slip \cite{duffy98}.
The resulting channel is 3~\unit{cm}, 30~\micron deep
and 1~\unit{mm} wide.
Spheres travel along the channel in a parabolic Poiseuille
flow profile, their speed depending on their height in 
the channel.

The flowing spheres interact with an array of
holographic optical traps \cite{HOT,polin05}
whose focused intensity pattern is recorded
in Fig.~\ref{fig:schematic}(b).
The pattern of 103 optical tweezers \cite{ashkin86} is created from a single
laser beam by imprinting
a computer-generated hologram \cite{polin05} onto its wavefronts
using a phase-only spatial light modulator (SLM; Hamamatsu X8267-16).
The hologram is relayed 
to the objective lens (Nikon Plan-Apo, 100$\times$, NA 1.4, oil immersion) 
of an inverted optical microscope (Nikon TE-2000U),
which focuses it into diffraction-limited traps
at the midplane of the sample \cite{roichman06c}.
Figure~\ref{fig:schematic}(b)
was created by placing a front-surface mirror in the 
microscope's focal plane, collecting the reflected light 
with the objective lens,
and recording it with the microscope's video camera
(NEC TI-324A II).
The traps are powered by a solid
state laser (Coherent Verdi 5W) operating at a vacuum
wavelength of $\lambda = 532~\unit{nm}$.

The optical tweezer array
was designed to sort an input distribution of colloidal spheres
into two fractions that are kinetically locked in to two distinct
lattice directions, and thus are deflected by the array in opposite
directions.
The array consists of two oblique lattice domains whose
$[10]$ axes are oppositely inclined at
$\theta_{[10]} = \pm 30^\circ$ with respect to the flow direction.
Photometry performed on the traps' images suggests that
their intensities vary by roughly 10\%.
This is consistent with calibration measurements based on the
thermally-driven fluctuations of spheres trapped in each tweezer
under zero-flow conditions \cite{polin05,roichman06a}.
Trajectories that become kinetically locked in to the $[10]$ direction,
such as the two measured examples denoted 
by circles in Fig.~\ref{fig:schematic}(b),
are deflected toward the apex of the array before flowing downstream.
Other particles that interact too weakly with the traps to be
deflected by $30^\circ$ may still become locked in to the $[11]$
direction, which is inclined at $\theta_{[11]} = \mp 13^\circ$
in the two domains.
Particles traveling along this direction are deflected away from
the apex to the sides of the array,
as indicated by the two typical trajectories denoted by
squares in Fig.~\ref{fig:schematic}(b).
The static array of traps therefore sorts the sample
into two distinct fractions, one 
concentrated at the
apex of the array, and the other deflected to its sides.

The objective lens
also is used to holographically 
track and characterize individual spheres as they
travel through the array.
To do this, the sample volume is illuminated
with the collimated 3~\unit{mm} diameter
beam from a HeNe laser operating at a 
vacuum wavelength of 632.8~\unit{nm}.
The laser's irradiance is on the order of $1~\unit{nW/\micron^2}$,
which is less than $10^{-4}$ of an individual trap's irradiance
and is comparable to that of a conventional incandescent illuminator.
Light scattered by a sphere interferes with
the unscattered portion
of the beam to form an interference pattern in the microscope's focal plane.
This is magnified and its intensity recorded
by the video camera with a calibrated
spatial resolution of 91~\unit{nm/pixel} at 30~\unit{frames/s}.
Figure~\ref{fig:schematic}(c) shows a detail of a
typical holographic snapshot
of 1.5~\micron diameter silica spheres
normalized by a background image to eliminate speckle and other static
nonuniformities \cite{LMM}.

Each snapshot is analyzed \cite{LMM}
using the Lorenz-Mie theory of light scattering \cite{bohren83}
to obtain each colloidal sphere's three-dimensional position,
$\vec{r}_p(t)$, relative to the
center of the focal plane at time $t$, its radius $a_p$, and its complex index
of refraction, $n_p$.
Rigorous error estimates \cite{LMM}
yield nanometer
resolution for $\vec{r}_p$ and $a_p$, and
part-per-thousand resolution for $n_p$.
Such precise single-particle characterization makes possible a
direct test of optical fractionation's resolution.

\begin{figure*}[!t]
  \centering
  \includegraphics[width=\textwidth]{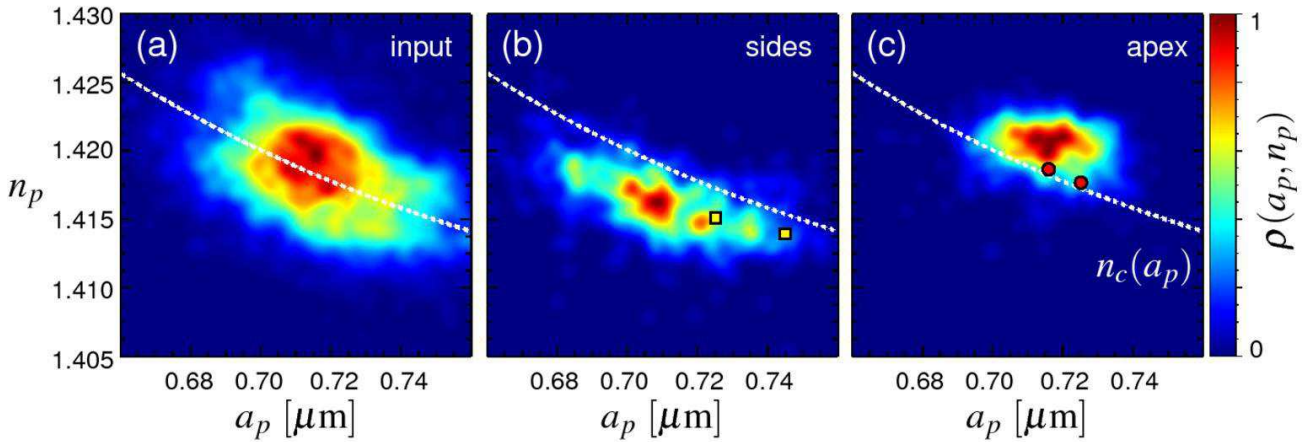}
  \caption{(color online) High-resolution optical fractionation of 
    monodisperse colloidal silica spheres demonstrated in the measured
    probability density $\rho(a_p,n_p)$ of particles with radius $a_p$
    and refractive index $n_p$.  The input sample, whose distribution
    is shown in (a), is
    separated into a smaller, low-index fraction in the side
    regions (b) and a larger, high-index fraction at the apex (c).  The
    dashed curves show the marginally locked-in condition
    $n_c(a_p)$ for the array's $[10]$ direction.
    Plotted squares and circles indicate the properties of the
    four representative spheres whose trajectories are plotted
    in Fig.~\protect\ref{fig:schematic}(b).}
  \label{fig:monodisperse}
\end{figure*}

A particle becomes kinetically locked-in to one of the array's
crystallographic directions
because its interaction
with one trap deflects it toward the
next trap along that direction, and so on through the lattice
\cite{korda02b,gopinathan04,ladavac04,pelton04a,gleeson06,herrmann09}.
Modeling an individual trap as a Gaussian potential
well of width $\sigma_p$ and depth $V_p$ separated from its neighbors
by a distance $b$, the maximum angle $\theta_\text{max}$
by which a particle driven by force $\vec{F}_p$ can be
deflected has been estimated
\cite{ladavac04,pelton04a,roichman07a} to satisfy
\begin{equation}
  \label{eq:lockinangle}
  \theta_\text{max} 
  \leq 
  \sin^{-1} \left(
    \frac{2}{\sqrt{e}} \, 
    \frac{V_p}{\sigma_p \, F_p} \,
    \exp\left( - \frac{b^2}{8 \sigma_p^2} \right) \right).
\end{equation}
This is a general result that applies equally well
to arrays of potential barriers 
as to potential wells \cite{pelton04a,herrmann09}.
In the particular case of a micrometer-scale colloidal sphere
interacting with optical tweezers, the effective well
depth depends on both the particle's size and refractive index,
and is given approximately by
\cite{harada96c,pelton04a},
\begin{equation}
  \label{eq:rayleighwell}
   V_p = \frac{n_m a_p^3}{\sigma_p^2 \, c} \, 
   \left( \frac{n_p^2 - n_m^2}{n_p^2 + 2 n_m^2} \right) \, P,
\end{equation}
where $n_m$ is the refractive index of the medium and
$P$ is the laser power in an individual trap.
The well's effective width \cite{ladavac04,pelton04a},
$\sigma_p^2 = a_p^2 + \lambda^2/(4 n_m^2)$,
also depends on the particle's size because larger spheres
encounter a trap's localized light field from greater distances.
Finally, the driving force due to Stokes drag in a fluid of viscosity $\eta$
flowing at velocity $\vec{v}$ is
\begin{equation}
  \label{eq:stokesdrag}
  \vec{F}_p = 6 \pi \eta a_p \, \vec{v},
\end{equation}
assuming that hydrodynamic coupling to boundaries and other spheres
is weak enough to ignore \cite{happel91}.

Equations~(\ref{eq:lockinangle}), (\ref{eq:rayleighwell}) and
(\ref{eq:stokesdrag})
define conditions for marginally locked-in transport
that divide the $(a_p,n_p)$ plane into a class of particles
that can be deflected to angle $\theta$ and a complementary class
that cannot.
This division falls along the curve
\begin{equation}
  n_c(a_p) = n_m \, \sqrt{\frac{A(a_p) + 2B}{A(a_p) - B}}, 
  \quad \text{where}
  \label{eq:lockincondition}
\end{equation}
\begin{equation}
  A(a_p) = 
  \left(\frac{a_p}{b}\right)^2 \,
  \left(\frac{b}{\sigma_p}\right)^3 \,
  \exp\left(- \frac{b^2}{8\sigma_p^2}\right)
\end{equation}
depends on particle size through the ratio $a_p/b$ and
\begin{equation}
  B = \frac{\sqrt{e}}{2} \, \frac{6\pi \eta \, v \, c}{n_m P}
  \, b \sin \theta
\end{equation}
depends only on control parameters.
Particles satisfying $n_p \ge n_c(a_p)$ should become locked in to the
lattice direction at angle $\theta$, and
those with $n_p < n_c(a_p)$ should escape,
perhaps to become locked in along other lattice directions
at smaller angles.
This result provides a basis to tune the selectivity
of optical fractionation for sorting by 
size and refractive index.

The data in Fig.~\ref{fig:monodisperse}
show results obtained with a single sample of 
monodisperse silica spheres
(PolySciences Catalog \#24327, Lot \#600424).
Lorenz-Mie characterization of 6000 randomly chosen spheres
yields a sample-averaged radius of $a_p = 0.715 \pm 0.021~\micron$,
which is significantly smaller and more monodisperse than
the manufacturer's specification of $0.75 \pm 0.04~\micron$.
The sample-averaged refractive index of 
$n_p = 1.418 \pm 0.004$ 
is consistent with typical values for colloidal silica spheres \cite{LMM}.
The distribution of input particle properties in
Fig.~\ref{fig:monodisperse}(a) was computed from single-particle
results using an 
optimal non-parametric density estimator \cite{silverman92}.
It reveals an anticorrelation between particle size and refractive
index that is typical for porous 
colloidal spheres grown by emulsion polymerization
and is not observed in fully compact particles
such as fluid droplets \cite{LMM}.

The homogeneous input sample was flowed at 
$v = 34 \pm 2~\unit{\micron/s}$ 
through the array of traps.
This is slow enough that blurring during the camera's
1~\unit{ms} exposure time does not degrade
holographic particle tracking and characterization \cite{LMM}.
Although particles approach the array at a range of heights
and thus at a variety of speeds, they are drawn to the channel's 
midplane by optical forces, and thus pass through at a uniform
speed.
This vertical flow focusing is made apparent by the three-dimensional
tracking capabilities of holographic video microscopy.
The concentration of particles is low enough that no more than
eight occupy the array at any time.
Collisions are rare at this low occupancy and no effort was made
to exclude their effects from our results.

The trap array's $[10]$ lattice spacing of $b = 2.03~\micron$ was
chosen so that roughly half of the sample would be kinetically locked-in
according to Eq.~(\ref{eq:lockincondition}) with a laser power 
of $P = 3~\unit{mW/trap}$.
The critical condition for locked-in transport along the $[10]$
lattice direction predicted by
Eq.~(\ref{eq:lockincondition}) is plotted as a dashed curve in
Fig.~\ref{fig:monodisperse} with no adjustable parameters.
The $[01]$ lattice spacing of $a = 3.27~\micron$ was selected
to be large enough to prevent locked-in particles 
from escaping \cite{pelton04a}, but small enough that the
low-angle $[11]$ direction would be able to
entrain and deflect all of the particles in the sample.
The array thus preferentially deflects particles
toward its apex, but provides an alternative route
with opposite deflection
for particles that interact too weakly to lock in to its
principal axis.
The two fractions thus are spatially separated from each other and also
from those particles that bypass the array's catchment basin altogether.

The characteristics of particles flowing into and out of 
the array were assessed within the rectangular 
regions indicated in Fig.~\ref{fig:schematic}(b),
with results from the two side regions being combined.
Particle-by-particle characterization in the side and apex regions,
plotted in Figs.~\ref{fig:monodisperse}(b) and (c),
reveals clear separation into
two distinct and sharply defined sub-populations, the larger, higher-index
particles being deflected to the apex and the smaller, lower-index
particles being deflected to the sides, in excellent quantitative
agreement with Eq.~(\ref{eq:lockincondition}).
The mean radius of the focused population is slightly
larger than that of the other fraction
($0.716 \pm 0.021~\unit{\micron}$ versus $0.710 \pm 0.027~\unit{\micron}$).
A far more substantial distinction is observed in the refractive
indexes of the sorted populations.
The focused spheres have a mean refractive index of
$1.420 \pm 0.003$ whereas the dispersed fraction
has a mean refractive index of $1.416 \pm 0.003$.
The properties of the
two spatially separated populations differ with better than
99.99 percent confidence according to the Wilcoxon rank-sum test.

The results in Fig.~\ref{fig:monodisperse} 
demonstrate that optical fractionation
is inherently a multidimensional sorting technique
whose selectivity for size and refractive index can be
dynamically optimized by tuning the properties of the optical trapping pattern.
The present realization demonstrates
discrimination on the basis
of refractive index with part-per-thousand resolution
under the set of conditions that we considered.
Similarly good results, including agreement with
Eq.~(\ref{eq:lockincondition}), were obtained for a monodisperse
sample of micrometer-diameter polystyrene spheres.
These observations provides experimental support for 
the suggestion \cite{korda02thesis,gopinathan04,pelton04a}
that the large number of particle-trap interactions in a typical
trajectory renders kinetic lock-in robust against
the noise and quenched disorder that arise in practical implementations.
A rigorous explanation for this robustness remains an outstanding
challenge, as does the quantitative predictive power 
of Eq.~(\ref{eq:lockincondition}), which was derived with
semi-quantitative limiting arguments \cite{ladavac04,pelton04a}.

The inhomogeneous trapping pattern used for this study
implements a prismatic mode of optical fractionation 
\cite{korda02b,gopinathan04} in which
different fractions become kinetically locked in to different
crystallographic directions and thus are sorted into spatially
separated streams.
More complex trapping patterns should enable still more sophisticated
multi-dimensional sorting with similarly large displacements for
finely resolved populations.
The very high-resolution fractionation by refractive index that
we have demonstrated should have immediate
applications in photonics, where colloidal particles' optical properties
increasingly are being exploited, and also in medical diagnostics, where
pre-selecting probe particles for precise optical properties may
reduce the time, effort and expense of bead-based molecular binding assays
\cite{LMM}.
The throughput for such applications can be increased by proportionately
increasing the flow speed and laser power.

This work was supported by the MRSEC program of the National
Science Foundation through Grant Number DMR-0820341.


\end{document}